\documentclass[number,preprint,review,12pt]{elsarticle}
\usepackage{hyperref}

\journal{Journal of Network and Computer Applications}

\begin{document}

\begin{frontmatter}

%% Title, authors and addresses %TODO

\title{Hybrid Cloud Architectures for Research Computing: Applications and Use Cases}

%\title{Hybrid Cloud Architectures for Research Computing in Life Sciences: Applications and Use Cases}

%\tnotetext[t1]{A title note to aknowledge supporters.}
% Author contributions: CRediT. Corresponding authors are required to acknowledge co-author contributions using CRediT (Contributor Roles Taxonomy) roles:

%% Affiliations
\affiliation[cmg]{organization={AG Computational Metagenomics, Faculty of Technology, Bielefeld University},
            city={Bielefeld},
            postcode={33615},
            country={Germany}}

\affiliation[unibas]{organization={Computational and Systems Biology, Biozentrum, University of Basel},
            city={Basel},
            postcode={4056},
            country={Switzerland}}

\affiliation[sib]{organization={Swiss Institute of Bioinformatics, Biozentrum, University of Basel},
            city={Basel},
            postcode={4056},
            country={Switzerland}}

\affiliation[ibg5]{organization={IBG-5, Forschungszentrum Jülich GmbH},
            city={Jülich},
            postcode={52428},
            country={Germany}}

\affiliation[ebi]{organization={EMBL-EBI, Wellcome Genome Campus},
            city={Cambridge},
            postcode={CB10 1SD},
            country={UK}}

\affiliation[ics]{organization={Institute of Computer Science, Masaryk University},
            addressline={Šumavská 525/33},
            city={Brno},
            postcode={60200},
            country={Czech Republic}}

\affiliation[elixir]{organization={ELIXIR Hub, Wellcome Genome Campus},
            addressline={},
            city={Cambridge},
            postcode={CB10 1SD},
            country={UK}}

\affiliation[unipd]{organization={Department of Biomedical Sciences, University of Padova},
            addressline={Via Ugo Bassi 58/B},
            city={Padova},
            postcode={35131},
            country={Italy}}

\affiliation[csc]{organization={CSC – IT Center for Science Ltd.},
            addressline={P.O Box 405},
            city={Espoo},
            postcode={02101},
            country={Finland}}

\affiliation[charite]{organization={Berlin Institute of Health at Charité - Universitätsmedizin Berlin},
            addressline={Luisenstraße 65},
            city={Berlin},
            postcode={10117},
            country={Germany}}           

%% Authors
\author[cmg]{Xaver Stiensmeier}
\ead{xaver.stiensmeier@uni-bielefeld.de}
\ead[url]{https://orcid.org/0009-0005-3274-122X}

\author[unibas,sib]{Alexander Kanitz}
\ead{alexander.kanitz@unibas.ch}
\ead[url]{https://orcid.org/0000-0002-3468-0652}

\author[ibg5]{Jan Krüger}
\ead{j.krueger@fz-juelich.de}
\ead[url]{https://orcid.org/0009-0006-4245-1653}

\author[ebi]{Santiago Insua}
\ead{sinsua@ebi.ac.uk}
\ead[url]{https://orcid.org/0000-0002-9247-243X }

\author[ics]{Adrián Rošinec}
\ead{adrian@ics.muni.cz}
\ead[url]{https://orcid.org/0000-0002-7748-5590}

\author[ics]{Viktória Spišaková}
\ead{spisakova@ics.muni.cz}

\author[ics]{Lukáš Hejtmánek}
\ead{Hejtmanek@ics.muni.cz}
\ead[url]{https://orcid.org/0000-0002-2078-8638}

\author[ebi]{David Yuan}
\ead{davidyuan@ebi.ac.uk}
\ead[url]{https://orcid.org/0000-0003-1075-1628}

\author[elixir,unipd]{Gavin Farrell}
\ead{gavin.farrell@elixir-europe.org}
\ead[url]{https://orcid.org/0000-0001-5166-8551}

\author[elixir]{Jonathan Tedds}
\ead{jonathan.tedds@elixir-europe.org}
\ead[url]{https://orcid.org/0000-0003-2829-4584}

\author[csc]{Juha Törnroos}
\ead{juha.tornroos@csc.fi}
\ead[url]{https://orcid.org/0000-0001-9216-0455}

\author[charite]{Harald Wagener}
\ead{harald.wagener@bih-charite.de}
\ead[url]{https://orcid.org/0000-0003-1073-4991}

\author[cmg,ibg5]{Alexander Sczyrba}
\ead{a.sczyrba@fz-juelich.de}
\ead[url]{https://orcid.org/0000-0002-4405-3847}

\author[ibg5]{Nils Hoffmann\corref{cor1}}
\ead{n.hoffmann@fz-juelich.de}
\ead[url]{https://orcid.org/0000-0002-6540-6875}

\author[ics]{Matej Antol\corref{cor1}}
\ead{antol@muni.cz}
\ead[url]{https://orcid.org/0000-0002-1380-5647}

%% Corresponding Author
\cortext[cor1]{Corresponding authors.}

%% Abstract
\begin{abstract}
Scientific research increasingly depends on robust and scalable IT infrastructures to support complex computational workflows. With the proliferation of services provided by research infrastructures, NRENs, and commercial cloud providers, researchers must navigate a fragmented ecosystem of computing environments, balancing performance, cost, scalability, and accessibility. Hybrid cloud architectures offer a compelling solution by integrating multiple computing environments to enhance flexibility, resource efficiency, and access to specialised hardware.

This paper provides a comprehensive overview of hybrid cloud deployment models, focusing on grid and cloud platforms (OpenPBS, SLURM, OpenStack, Kubernetes) and workflow management tools (Nextflow, Snakemake, CWL). We explore strategies for federated computing, multi-cloud orchestration, and workload scheduling, addressing key challenges such as interoperability, data security, reproducibility, and network performance. Drawing on implementations from life sciences, as coordinated by the ELIXIR Compute Platform and their integration into a wider EOSC context, we propose a roadmap for accelerating hybrid cloud adoption in research computing, emphasising governance frameworks and technical solutions that can drive sustainable and scalable infrastructure development.
\end{abstract}


\begin{highlights}
\item We identify the challenges of navigating a fragmented ecosystem of computing environments between research infrastructures, NRENs, and commercial clouds, and position hybrid cloud architectures as a solution to balance performance, cost, scalability, and accessibility.
 
\item We describe deployment models and workflow tools to enable federated computing, multi-cloud orchestration, and workload scheduling, while tackling key challenges like interoperability, security, and reproducibility.
 
\item We demonstrate practical implementation through functional demonstrators developed within the ELIXIR Compute Platform work program and propose a roadmap for hybrid cloud adoption that prioritizes governance frameworks and sustainable infrastructure development for research.
\end{highlights}

%% Keywords
\begin{keyword}
Hybrid Cloud \sep Multi-Cloud \sep Workflow Management \sep Research Infrastructure  \sep EOSC \sep ELIXIR \sep Federated Computing % TODO Only 1-7 are allowed; drop 3
% \sep HPC \sep Data Interoperability \sep Life Sciences




\end{keyword}

\end{frontmatter}


Research communities across different scientific domains require increasingly robust and scalable IT infrastructure to support their projects. These requirements can span from enormous computational and storage capacities to more specialised applications harnessing computing resources such as GPUs. To address such infrastructure requirements, computing resources are continually being built on a national and international level within ESFRI research infrastructures, NRENs and other similar facilities. Additionally, commercial IT service providers have started to offer a broader spectrum of services that provide solutions to support research activities.

To maximise these diverse computational infrastructures and optimise their utilisation, researchers and their supporting organisations must continually deploy their computational workflows in the most suitable environments. However, moving between different computing environments instead of staying with one that may be sub-optimal requires extra time and effort as well as technical expertise from researchers to plan and execute such a migration.

These challenges can be partially addressed by leveraging hybrid cloud architectures, the concept of seamlessly combining multiple different cloud environments. Such multi-environmental deployment promises to increase scalability, expand available functionality, decrease costs and provide access to special resources such as GPUs.

Across research use cases, the most prominent drivers to utilise hybrid cloud architectures are scalability, access to specialised compute resources and increasingly, access to sensitive data \cite{EOSCTITANProjecta}, \cite{EuropeanHealthData2024a}, \cite{EuropeanNetworkTrusteda}.

In this paper, we provide novel insights on how researchers can harness multiple computing environments and apply diverse hybrid cloud deployment models for executing scientific computing workflows. Specifically, we focus on grid and cloud systems built on OpenPBS \cite{OpenPBSOpenSourcea}, SLURM \cite{SlurmWorkloadManager2023a}, OpenStack \cite{OpenStack2023a} and Kubernetes \cite{Kubernetes2023a}. We explore several use case scenarios using the workflow managers NextFlow \cite{ditommasoNextflowEnablesReproducible2017b}, Snakemake \cite{kosterSnakemakeScalableBioinformatics2012} and CWL \cite{crusoeMethodsIncludedStandardizing2022a}. These environments and workflow managers are used by researchers across scientific domains \cite{wrattenReproducibleScalableShareable2021a}, although our demonstrators focus on use cases from the life science domain specifically.

This paper can be viewed as a stepping stone toward reducing the obstacles faced by researchers looking to optimize their use of computing power via research infrastructures within and beyond the ELIXIR Europe and EOSC ecosystems \cite{commissionRealisingEuropeanOpen2016}. More specifically, it outlines different approaches to provide seamless access to a combination of multiple computing and storage environments within single computational pipelines.

\subsection*{Terms and Abbreviations} % TODO separate document Glossary: Please provide definitions of field-specific terms used in your article, in a separate list.

\begin{description} % [labelwidth=1.7cm] % (to align all; leads to more linebreaks)
  \item[\textbf{ASIC}] Application Specific Integrated Circuit
  \item[\textbf{EOSC}] European Open Science Cloud
  \item[\textbf{EHDS}] European Health Data Space
  \item[\textbf{ELIXIR}] European Life Sciences Infrastructure
  \item[\textbf{ESFRI}] European Strategy Forum on Research Infrastructures
  \item[\textbf{FAIR data}] Findable, Accessible, Interoperable \& Reusable data
  \item[\textbf{FPGA}] Field Programmable Gate Array
  \item[\textbf{GAIA-X}] European Association for Data and Cloud AISBL
  \item[\textbf{GDI}] Genomics Data Infrastructure
  \item[\textbf{GDPR}] General Data Protection Regulation
  \item[\textbf{GÉANT}] Pan-European data network for the research and education community
  \item[\textbf{GPU}] Graphics Processing Unit
  \item[\textbf{HPC}] High-performance Computing
  \item[\textbf{IaaS}] Infrastructure-as-a-Service
  \item[\textbf{ISO}] International Standards Organization
  \item[\textbf{FaaS}] Function-as-a-Service
  \item[\textbf{CaaS}] Container-as-a-Service
  \item[\textbf{SaaS}] Software-as-a-Service
  \item[\textbf{NREN}] National Research and Education Network
  \item[\textbf{OCRE}] Open Clouds for Research Environments
  \item[\textbf{OS}] Operating System
  \item[\textbf{OpenPBS}] Portable Batch System - Scheduling of HPC jobs
  \item[\textbf{SLURM}] Simple Linux Utility for Resource Management - Scheduling of HPC jobs
  \item[\textbf{SPU}] Stream Processing Unit
\end{description}

\section{The Evolving Landscape Of Research Computing: From Grids to Clouds and Beyond}

\subsection{Infrastructures and services for research}
\label{s:inf_and_serv_for_res}
Scientific communities form a large vibrant ecosystem, centred around universities, research institutes and research infrastructures. While universities and research institutes provide a fertile ground for scientists to focus on research and education, research infrastructures create networks of scientists within a specific field across and beyond countries, providing the basis for cooperation and creating space for bootstrapping domain-specific innovation. Research infrastructures can thus provide an environment fostering collaboration, innovation and service delivery regardless of national and university affiliation.

International cooperation between research infrastructures, scientific communities and other facilities has taken place for decades. In Europe, the so-called ESFRI\footnote{European Strategy Forum on Research Infrastructures} recognises more than 60 research infrastructures with pan-European relevance \cite{infrastructuresESFRIRoadmap20212021}. Next to research infrastructures, another environment for supporting research emerged within European NRENs, which share their common expertise in advancing research, education and innovation within a single association called GÉANT \cite{GEANTAnnualReporta} and the EGI foundation which operates and provides compute infrastructure for individual researchers and research infrastructures \cite{EGIAdvancedComputinga}. All these and other infrastructures and other institutions clustering researchers across scientific domains continuously develop and maintain robust environments for storing, processing and analysing data, which lie at the heart of modern research endeavours.

The cooperation within and between these infrastructures and institutions has lately been reinforced by efforts encompassed by EOSC, which seeks to provide European researchers, innovators, companies and citizens with a federated and open multi-disciplinary environment where they can publish, find and re-use data, tools and services for research, innovation and educational purposes. This data should adhere to the FAIR principles \cite{wilkinsonFAIRGuidingPrinciples2016a}, i.e. follow the approaches which make them more Findable, Accessible, Interoperable and Reproducible. EOSC thus encompasses not only data management but also data processing, as the value of scientific data lies in their analysis and application to help solve research questions.

\subsection{Computing in academia – grids, supercomputers and clouds}
\label{ss:computing_in_academia}
Current state-of-the-art environments for processing of high volumes of scientific data conventionally utilise batch processing \cite{iso/iec2382:2015enISOIEC23822015a}, which is usually designed as a distributed system consisting of multiple sites with computing resources. Centralised systems used in high-performance computing (HPC) are supercomputers consisting of highly specialised hardware for computing, networking and storage. They share the idea of a grid system in terms of access to resources and overall architecture, but unite the hardware and physical location of computing resources in order to minimise latency and to optimise throughput, often with accelerator hardware like graphics or stream processing units (GPUs, SPUs), neural and tensor processing units (NPUs, TPUs) or bespoke application-specific integrated circuits (ASICs). The grid systems, owing to their specialization, have proven to often be too rigid for more complex scenarios, which has created space for emerging cloud technologies. These are getting more popular as they are dismissing a number of grid-specific disadvantages, as well as offering new opportunities in distributed computing approaches.

\begin{figure}
    \centering
    \includegraphics[width=0.8\linewidth]{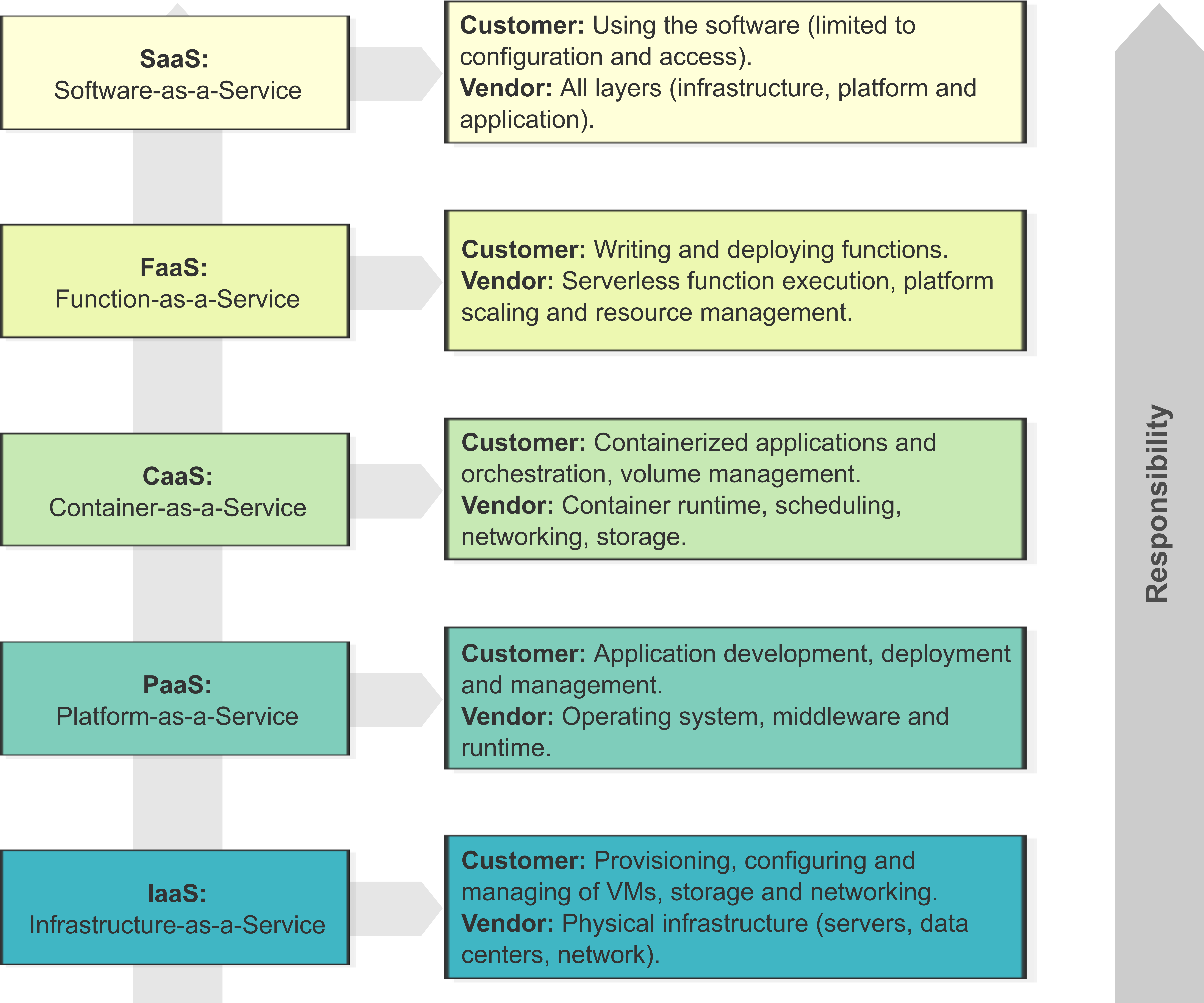}
    \caption{Service type abstractions from bottom to top and associated level of responsibility between customer and vendor concerning implementation and operation. IaaS: Infrastructure-as-a-Service; PaaS: Platform-as-a-Service; CaaS: Container-as-as-Service; SaaS: Software-as-a-Service; FaaS: Function-as-a-Service. From bottom to top, more responsibility concerning development, provisioning and operations are shifted from the customer towards the vendor.}
    \label{fig:spiaas}
\end{figure}

\textbf{In cloud computing} \cite{iso/iec22123-1:2023eInformationTechnologyCloud2023a}, service models provide a comprehensive framework of abstractions regarding the provisioning of virtualized and dedicated computing resources. The service model's primary role is to abstract the control level and delegate responsibility. On the lowest level, infrastructure-as-a-service (IaaS), users manage fully dedicated hardware resources and managed services of operating systems (OS) and consecutive layers by themselves. On the up-most layer, software-as-a-service (SaaS), users only interact with the microservice architecture and build their applications from various services created, managed, and provided by the cloud provider (Figure \ref{fig:spiaas}). These extremes outline a scale of cloud service models, which have a number of benefits and drawbacks to consider. Recent developments, such as Container-as-a-service (CaaS), share aspects of PaaS and SaaS, while Function-as-a-Service (FaaS) provides an even higher abstraction, exposing individual, composable functions to the consumer or customer.

Today, grid and cloud environments complement each other in their utilisation for research purposes. Researchers must choose and weigh trade-offs of varying pricing models and costs, functionality, applications, availability and specialised support, physical closeness to resources and more. They also have to spend time and resources not only to select the best environment but even to map and compare all the available options in the first place. Last but not least, choosing a specific environment now predetermines further development and future investments, regardless of changes in both requirements and available solutions, often until the chosen environment is no longer physically or financially viable. The flexibility of containerised services deployed on virtualised commodity hardware and infrastructure have significantly reduced that burden. But still, migrating previous work and going through a learning process related to new technologies and services still remain costly in terms of time and training.

\subsection{Deployment models}
\label{ss:deployment_models}

The above-mentioned environments for data processing have different strengths and weaknesses regarding their utilisation, including but not restricted to scalability, service characteristics and functions. On top of that, regardless of restrictions on a technological level, different distributions and implementations of specific environments deal with necessary trade-offs in their configuration based on already known preferences, creating artificial obstacles where true innovation requires flexibility. It is thus only natural to consider architectures that seek to put all these resources together without unnecessary barriers, leveraging the benefits they may jointly provide.

Following the ISO/IEC 22123(2023) definitions \cite{iso/iec22123-1:2023eInformationTechnologyCloud2023a}, we will use the following deployment models for clouds:

While some of the cloud-infrastructures may be classified as \textit{public clouds}, especially ones offered by commercial providers, most cloud infrastructures operated for science are \textit{community clouds}. Depending on the operational model, these may consist of multiple sites that are operated by the same legal entity, or of multiple sites that are operated by independent legal entities, bound by common policies, processes and contracts as a federated cloud.\cite{hoffmannEmbeddingDeNBICloud2023a}

The cloud infrastructure of a hybrid cloud \cite{mellNISTDefinitionCloud2011a} is composed of two or more distinct cloud infrastructures, e.g., a public and a private part. Although these infrastructures are unique entities, they are connected by standardized and often proprietary technology. Thanks to this bonding, applications and data are portable between infrastructures. Connecting private and public infrastructures in a way that allows for portability of data and applications is beneficial when privacy concerns are present, as confidential data can be stored in private infrastructures.

A multicloud \cite{iso/iec22123-1:2023eInformationTechnologyCloud2023a} infrastructure utilises and allows the usage of two or more of the same deployment models (e.g., separate public clouds). The main advantage of this architecture is the improved reliability by creating a high-availability solution – in case of outage of cloud services of any provider being temporarily unavailable, applications are deployed in another provider's infrastructure as a passive or active fallback.

\subsection{Current landscape of compute technologies}
The combination of containerization, workflow managers, and common execution platforms lays the groundwork for managing diverse computational workloads. This allows organizations to benefit from streamlined operations and enhanced scalability.

% These paragraphs could be transformed to \subsubsections

\paragraph{Containerization}
Containerisation has become a cornerstone technology in modern cloud environments. Containers, such as those created with Docker \cite{DockerAcceleratedContainer2022a} or Apptainer (formerly: Singularity) \cite{ApptainerPortableReproduciblea} encapsulate applications and their dependencies, ensuring consistent execution across various environments. This technology simplifies application deployment and management while promoting scalability and portability. Containers are instrumental in ensuring that software components run reliably and reproducibly, making them a foundational building block for cloud-native applications.
Initiatives that promote the use of container technology in life science and bioinformatics, such as BioContainers \cite{daveigaleprevostBioContainersOpensourceCommunitydriven2017a}, have helped standardize container usage and provide quick access to tools. This approach addresses the challenges researchers face regarding the reproducibility, software dependencies, and portability of bioinformatics tools.

\paragraph{Workflow managers}
Workflow managers play a critical role in scientific computing and bioinformatics by providing a systematic approach to designing, executing, and managing complex computational workflows. They enable researchers to streamline their analysis processes, improve reproducibility, and enhance collaboration. Popular workflow managers include Apache Airflow, Nextflow, CWL, and Snakemake.

\paragraph{Common execution platforms}
Common execution platforms are natively available for multiple cloud platforms and even for specific technologies like Kubernetes \cite{ExecutorsNextflowDocumentationa}. Nextflow, for instance, supports the execution of tasks within a single workflow in different environments using multiple executors for job scheduling and execution, such as SLURM or Kubernetes.

\section{Challenges of Computing in Research}
\label{s:challenges}
In this chapter, we outline a brief selection of the main challenges encountered by scientists while working with computing environments.

\paragraph{Resource scarcity in terms of both quality and quantity} Resource scarcity is the greatest challenge. As a rule, there is an everlasting lack of state-of-the-art hardware and software in on-premises infrastructures, although they tend to be more accustomed to the needs of their communities. On top of that, due to limits of the shared pool of resources, users may encounter a lack of availability of present resources as well. Major public cloud providers are more flexible in this regard, and in general may guarantee higher availability of resources, although the services they provide are more general in nature, and require more resources dedicated to workflow and application management and development on the user side.

\paragraph{Integration and Interoperability}
Another issue is the integration and interoperability of functionalities from different environments. The deeper involvement of commercial cloud providers in creating services that could be used by the scientific community thanks to the GAIA-X or OCRE projects provides easy access to additional resources, but at the same time involves integration via a proprietary API that may lead to vendor-dependence. An example of such a service is Google LifeScience API or Microsoft Genomics. Although the benefit of using such services is usually a lower entry-barrier, because all necessary services are offered by one vendor and are therefore compatible within the same ecosystem, this is paid for by a loss of flexibility to switch to another service should the current one be discontinued or no longer be feasible due to legal or financial reasons.

\paragraph{Reliability and High-Availability} Ensuring high availability [ISO 3.13.7 "availability"] is crucial for specific computational tasks. The ability to execute jobs across separate environments enhances high availability, preventing significant disruptions in one environment from impacting others. Workload managers play a pivotal role by automatically rescheduling unsuccessful jobs to environments that remain operational.

Moreover, this cross-environment capability extends to disaster recovery scenarios. In the event that one cloud becomes inaccessible, data mirrored in alternative environments remains accessible, providing a resilient strategy for maintaining data availability.

\paragraph{Reproducibility}
It has been shown that scientists are strongly incentivised to publish more positive than negative results \cite{duyxScientificCitationsFavor2017a} and encouraged to repeatedly re-submit rejected papers \cite{tiokhinHonestSignalingAcademic2021a} without validating their methods and results. Consequently, an alarmingly high portion of results cannot be reproduced, verified, or confirmed in any way \cite{ioannidisWhyMostPublished2005a}, \cite{ioannidisHowMakeMore2014a}. Computing infrastructures should address these issues by supporting reproducibility principles \cite{ziemannFivePillarsComputational2023a}. The promising directions are containerisation services, software quality and assurance practices like continuous building and testing, versioning of the software and computing pipelines, as well as clear documentation of computing resources (bill of materials, e.g. processor models, GPU models, network hardware etc.) and software parameters.

The need to run different bioinformatics tools, some of them with heavy dependencies on common libraries and versions, is one of the main obstacles when changing the execution environment or when executing workloads across multi cloud or hybrid cloud environments. Container technologies like Docker and Singularity have helped to improve portability significantly by bundling software and dependencies , therefor allowing easier and platform-independent re-use of such software as part of scientific workflows.

\paragraph{Provenance}
Provenance tracking in hybrid cloud environments is particularly challenging due to the distributed nature of resources, necessitating efficient metadata management and the implementation of compatible provenance tracking tools, such as ProvONE \cite{bettiviaProvONE2022a} or Apache Atlas \cite{ApacheAtlasDataa}. These tools help in maintaining a comprehensive record of data lineage and computational history, essential for scientific integrity and validation.

Furthermore, the FAIR data principles present additional hurdles in hybrid cloud setups. Ensuring data findability and accessibility across varied cloud environments, each with its own security and access protocols, is a non-trivial task. The issue of interoperability is accentuated by the disparate data formats and systems prevalent in these environments. To enhance data reusability, researchers must focus on standardised data documentation practices and the adoption of open licences where feasible.

\paragraph{Data Volume and Mobility}
Managing vast data volumes and ensuring efficient mobility is critical across fields such as genomics, proteomics, and high-resolution medical imaging. Genomics projects produce terabytes of sequencing data, while proteomics and imaging studies generate complex, multi-dimensional datasets that demand scalable storage, robust processing, and rapid transfer between research centers and cloud environments. These transfers are complicated by network constraints, latency, and stringent security protocols, underscoring the need for distributed architectures and interoperable management strategies. By combining distributed data management, adaptive network strategies, and interoperable infrastructures, hybrid clouds effectively address issues such as network latency, bandwidth limitations, and vendor lock-in, thus delivering a more agile and resilient research environment.

\paragraph{Data Integrity, Sensitivity and Security}

An increasing number of research activities require scientists to consider the sensitive nature of data they process. In life sciences, sensitive data typically needs to be handled at the interfaces between medical, health and pharmaceutical research, which presents both legal and technological challenges. The legal and security issues related to research conducted on sensitive data must be addressed. Regulations and laws such as GDPR \cite{gdpr2016} are continuously developed on both national and international levels, presenting ever-growing demands and restrictions on how scientists may and may not collaborate on data sensitive in nature. 

The gap between requirements for collaboration and sensitive data handling is addressed by a number of projects and initiatives, spanning from the concept of European Health Data Spaces (EHDS) \cite{EuropeanHealthData2024a}, to specifically focused activities such as the Genomic Data Infrastructure (GDI) \cite{EuropeanGenomicDatab}.

Federated architectures are consequently used as a blueprint that facilitates interoperability and information sharing between autonomous, decentralised nodes – data remain within jurisdictional boundaries, while metadata are centralised and searchable. Hybrid- and multi- cloud architectures fall well within this category, and as such, may prove as a suitable model for distributed environments processing sensitive data.

\paragraph{Operational Costs and Environmental Impact}
The usage of cloud resources provided by commercial and academic providers is subject to both cost and $CO_{2}$ footprint optimisation. Transitioning to higher datacenter efficiency and renewable, carbon-neutral energy sources is highly incentivised within the EU's green deal agenda \cite{GreenCloudGreen2023a}. However, disproportionally rising energy and supplier prices may diminish these efficiency gains in operational costs. Thus, autoscaling features (provided by SLURM and Kubernetes) and preemptible or \emph{spot} instances are required to help reduce operational costs from the customer side, but their management is complex for users due to their ephemerality.

\section{Hybrid Cloud Architectures in Research}
\label{s:demonstrators}
In this chapter, we present five distinct architectural approaches to what could be understood as a hybrid cloud for research applications. We present these approaches using the intuitive concept of layers, beginning with the most basic and user-driven scenario and progressing through federation, task execution, and workflow management to a hybrid cloud built on the infrastructure layer. These approaches were explored as part of the ELIXIR Compute Platform's work program from 2022 to 2023.

For clarity, we present the use cases alongside the original motivations for their development because they were driven by specific service application scenarios and domains. However, the presented implementations may be adapted to fit more general requirements.

\subsection{Manual or Naive Approach}
\label{ss:manual_ap}

The manual approach to a hybrid/multi-cloud architecture involves splitting independent data into batches, manually distributing these batches across various clouds as inputs for containerised workflows, and consolidating results in a common storage service.

Each cloud uses a containerized workflow, which involves packaging and isolating the tools required for each step of the workflow using a container framework and a workflow manager to orchestrate data input/output and computation. In bioinformatics, a typical example is Docker with Nextflow running on a SLURM-compatible execution platform. The workflow defines the common storage for outputs.

To achieve a common execution environment that guarantees seamless execution for each cloud, it is recommended that an Infrastructure-as-Code approach be used with automation tools such as BiBiGrid or Terraform to improve infrastructure reproducibility.

\begin{figure}
    \centering
    \includegraphics[width=0.7\linewidth]{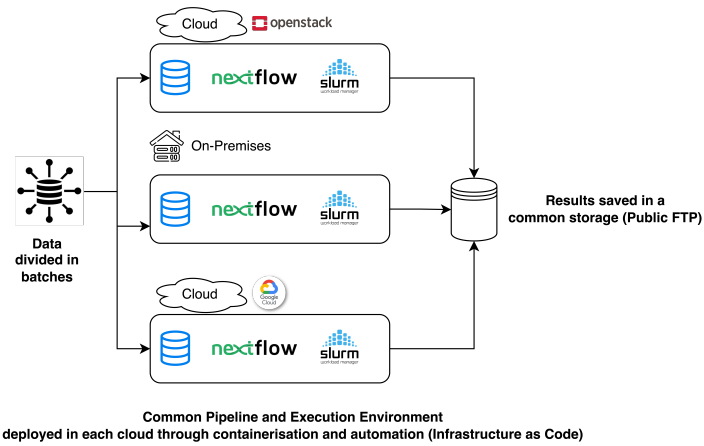}
    \caption{Deployment and execution flow for Nextflow and SLURM environments deployed independently to three different cloud or on-premise environments. Data flows from the source (left), is then split into batches that are then executed independently in one of the three environments, before result data is recombined in a common storage (right), FTP in this case.}
    \label{fig:manual_ap}
\end{figure}

In Figure \ref{fig:manual_ap} the manual approach is shown for OpenStack clouds with Nextflow workflows that execute on SLURM. Outputs are stored via FTP in a common storage. This approach is not practical if the data batches are interdependent, as several manual distributions of data batches would be required. The following approaches therefore automate the data distribution.

\subsection{Federated Computing and Data Staging with Centralised Metadata Repository}
\label{ss:federated_ap}
\begin{figure}
    \centering
    \includegraphics[width=0.7\linewidth]{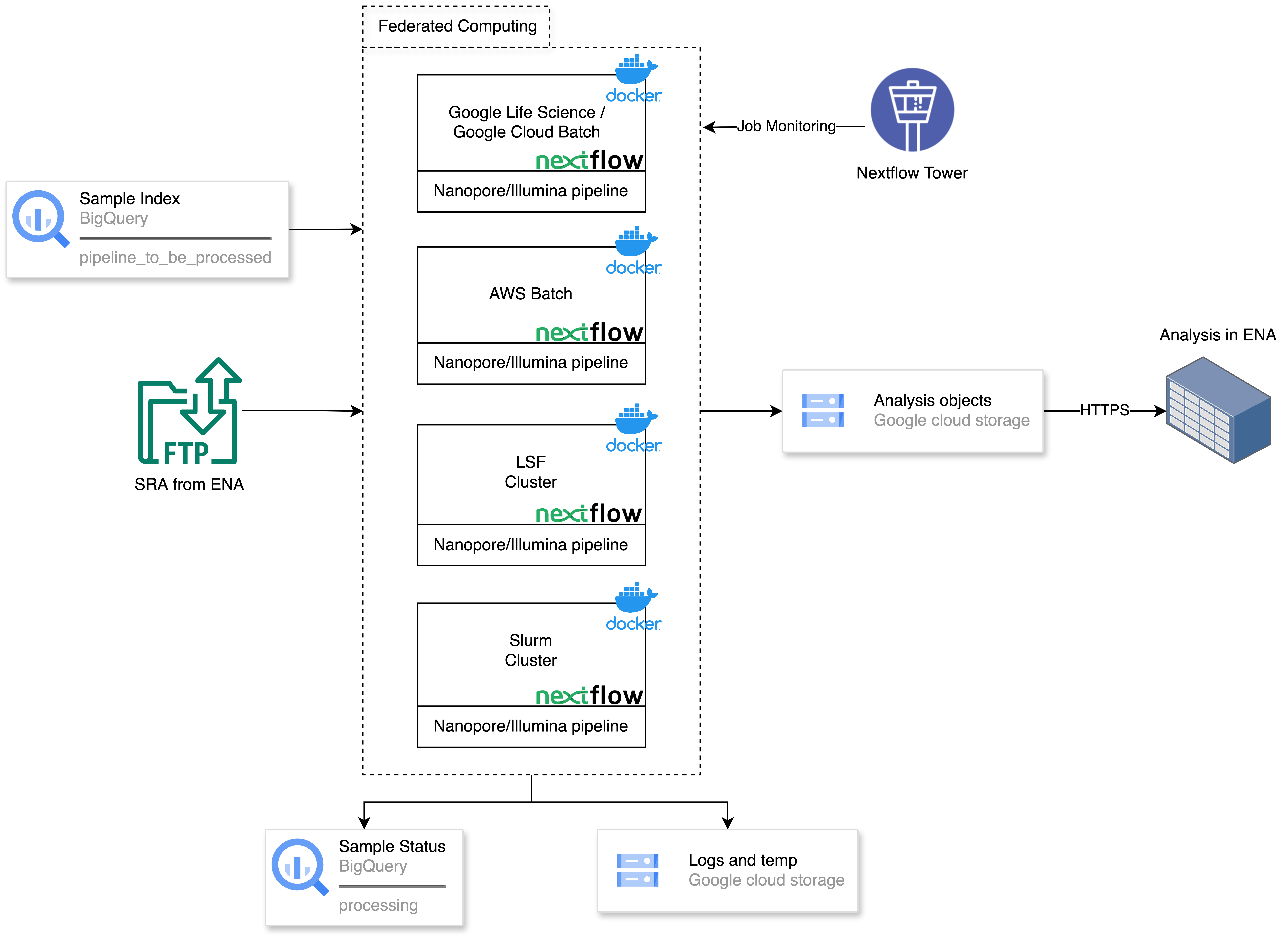}
    \caption{An implementation of the advanced solution in the second approach. The federated computing across multiple HPC clusters and multiple clouds (e.g. GCP \& AWS) was used for the systematic analysis of SARS-CoV-2 genomes in the COVID-19 pandemic.}
    \label{fig:federated_ap}
\end{figure}

To improve upon the general idea of the previous approach, the data batch distribution can be automated. The job scheduling across the hybrid or multi-cloud can also be made more versatile to efficiently handle constantly evolving and interdependent data \cite{sheffieldBiomedicalCloudPlatforms2022a}.
This second approach (Figure \ref{fig:federated_ap}) offers two solutions for these improvements. Both make use of a lightweight centralised metadata server that can be accessed by all independent clouds in order to receive information about what data to use. This data tracking system can follow the data journey from to-be-processed, through processing to succeeded or failed to output.\cite{yuanCovidsequenceanalysisworkflow2023} This is achieved by tagging the metadata. While a simple approach uses a central controller - a dominating workflow system - in order to schedule jobs to the workflow systems of the individual cloud, it also introduces a single point of failure and a performance bottleneck. A more advanced approach eliminates this central controller and works autonomously with the metadata repository.

\subsubsection*{Federated Execution Environment}
In this approach, each execution environment operates autonomously, interacting solely with the metadata repository. Execution environments reserve a batch of input data by tagging its metadata in the repository, perform analyses at its own pace and capacity and update the metadata again when the batch finishes. Then the metadata of the input are tagged with the status of success or failure accordingly. Together, all the federated execution environments work towards the common goal independently. Neither input nor output data need to be in the common storage, minimising expensive data transfer operations. Given the lightweight metadata repository, the federated execution environment model guarantees nearly unlimited scalability.

\subsection{Overlay networks and Multicloud over VPN infrastructure}
\label{ss:vpn_ap}
This approach focuses heavily on usability by abstraction through VPN and reproducibility realised by an automatic deployment via the cloud cluster creation and management tool BiBiGrid \cite{stiensmeierBiBiServBibigrid312025} which currently only supports OpenStack.

\begin{figure}
    \centering
    \includegraphics[width=0.6\linewidth]{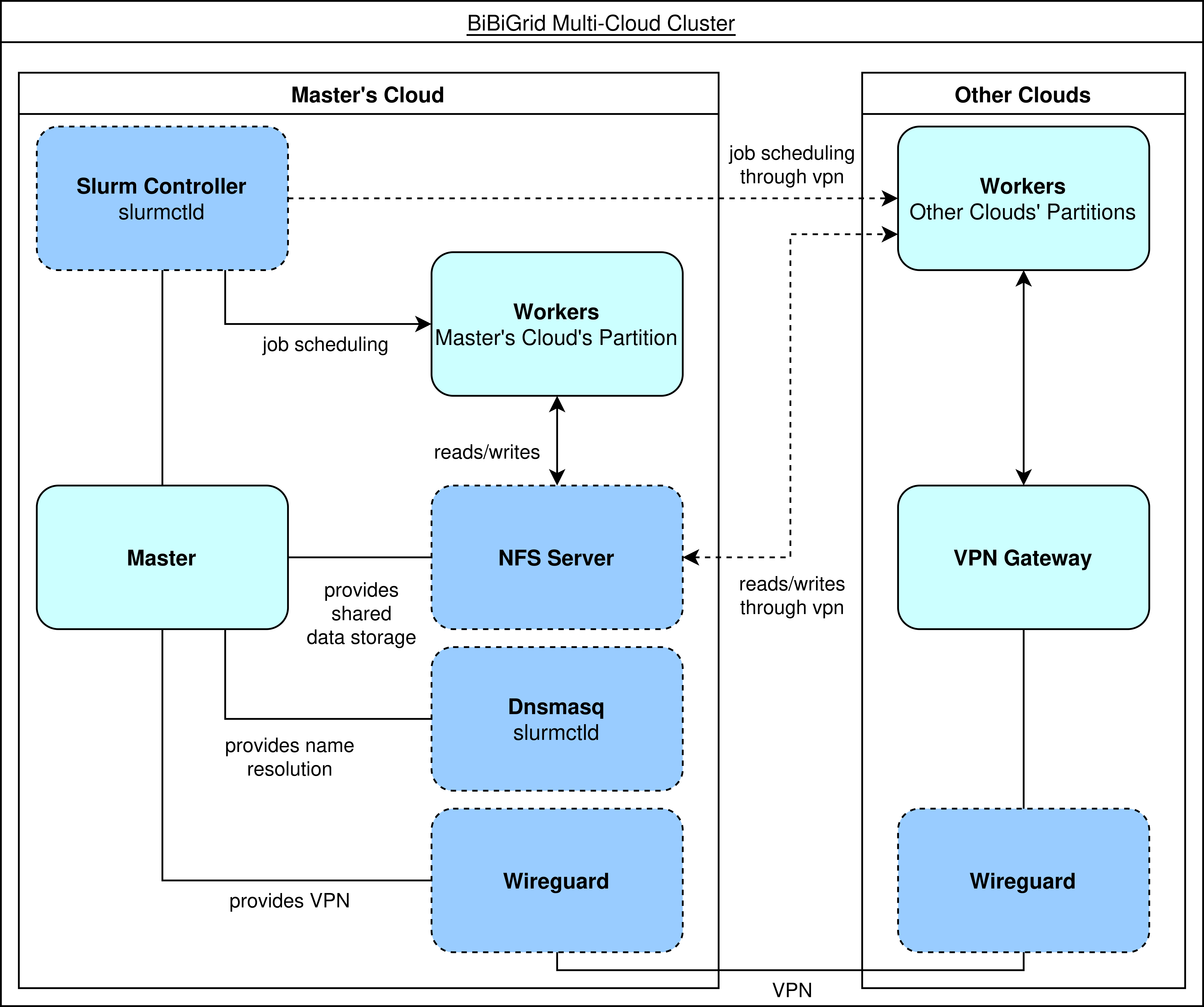}
    \caption{Schematic of a BiBiGrid Multi-Cloud Cluster. The main deployment, including the master server has control over local and remote job scheduling (the latter via VPN tunnel). Usage of Dnsmasq and Wireguard to set up the VPN allows the creation of a virtual  execution environment spanning multiple clouds. Data access for reading and writing is provided by a central NFS server, also accessible to the remote clouds via VPN.}
    \label{fig:vpn_ap}
\end{figure}

A single SLURM cluster is spanned across the entire hybrid- or multi-cloud deployment using Wireguard VPN \cite{donenfeldWireGuardFastModerna} with name resolution provided by Dnsmasq \cite{DnsmasqNetworkServicesa} which allows users to treat hybrid/multi-cloud as if it were a single cloud. The data is exchanged between all nodes in the VPN via NFS. This setup allows a single workflow system (e.g. Nextflow) to use the whole hybrid/multi-cloud by executing on the SLURM cluster, which then schedules jobs to any node within the VPN (Figure \ref{fig:vpn_ap}).

The user has complete control over job scheduling by specifying resource, feature, and partition requirements. BiBiGrid automatically creates a SLURM partition for each environment, allowing jobs to be scheduled to a specific cloud. Resource and feature requirements ensure that jobs are scheduled to the appropriate node.

Further development will focus on re-evaluating and improving SLURM's scheduling decisions in a hybrid/multi-cloud deployment scenario and in optimising data sharing between different clouds. The proposed hybrid/multi-cloud solution can be set up using the BiBiGrid Hybrid Cloud ELIXIR Hands-On tutorial \cite{BiBiGridHybridMultiCloud2024a}.

\subsection{Multicloud built on the workflow management layer}
\label{ss:wml_ap}
This section addresses the challenge of maximising resource utilisation in a hybrid cloud environment comprising a small but reliable Kubernetes cluster and a larger, albeit less reliable, OpenPBS cluster. The solution is centred around the development of an admission controller capable of transparently intercepting Kubernetes API calls and seamlessly routing them to the OpenPBS cluster when Kubernetes reaches maximum capacity.\cite{spisakovaCERITSCK8shpc2022}

Data management presents a significant challenge due to the lack of direct accessibility between storage systems linked to the Kubernetes and OpenPBS clusters. The job running in OpenPBS uses an sshfs mount for data input and output, computation is done in a scratch-like directory. Because the SSH proxy in our case must have a public IP address, and public IPs are generally a scarce resource, we use a single proxy to handle multiple jobs.

The operator was implemented using the Nextflow Sarek pipeline \cite{NFCoreSarekPipelinea}. The pipeline uses a native Kubernetes executor (kuberun) that runs every part of the pipeline in Kubernetes directly via the Kubernetes API. Some jobs from this pipeline that requested bigger resources can be moved to the OpenPBS cluster by our operator and the whole pipeline finished successfully.

\subsubsection*{Limitations}
The proposed operator has some limitations. First, only single node tasks are supported, i.e., it is not possible to move MPI based multi node jobs as there is no support for network connection between the Kubernetes and OpenPBS clusters. Second, only single container jobs are supported. This is mostly an implementation limitation than conceptual, however, singularity does not provide any means of sidecars, so probably co-located OpenPBS jobs would be needed in such a case. Third, data access via sshfs is not performance optimised and not suitable for production use, a better solution is needed. Finding a solution general enough to be deployed across different environments has proven challenging. S3 storage could be used as an alternative once it is widely adopted by applications and tools.

\subsection{Multicloud built on the task execution layer}
\label{ss:tel_ap}
The Global Alliance for Genomics and Health (GA4GH) \cite{GlobalAlianceGenomicsa} is developing standards for responsible data sharing in the life sciences. One such standard, the Task Execution Service (TES) API specification \cite{kanitzGA4GHTaskExecution2024a}, enables the execution of containerised workloads on various compute backends, thereby enabling hybrid and multi-cloud use cases.

The service ecosystem around the TES specification is growing, with several implementations for native cloud clusters, HPC/HTC clusters, and a well established global cloud service provider, a gateway TES implementation for injecting arbitrary middleware into TES requests, and client implementations, including TES-aware workflow engines like Snakemake \cite{molderSustainableDataAnalysis2021a}, Nextflow and CWL-TES.

To demonstrate the hybrid and multi cloud capabilities of the TES API, ELIXIR Cloud and AAI Driver Project of GA4GH is setting up a federated network of TES deployments in front of different compute cluster flavours. In this setup, the TES network sits behind a gateway with middleware that acts as a \textit{reverse proxy} to distribute incoming TES tasks (e.g., from a compatible workflow engine or another TES client) across the network according to rules that minimise data transfer. Specifically, each compute job is sent to the TES node that is physically situated closest to the input data. Jobs are computed and outputs are written to a centralised S3-based cloud storage (Figure \ref{fig:tel_ap}). Based on a simple Snakemake workflow with a scatter-gather step, a demonstrator has been developed to showcase the distribution of tasks across this network \cite{schneider-lunitzElixircloudaaiDemoteshybridcloud1002025}.

The TES community is actively working on addressing existing limitations (e.g., lack of standardised mechanism to pass credentials for storage access) by improving the standard, upgrading the services in the TES ecosystem and implementing more sophisticated, real-world use cases. The ELIXIR Compute Platform's new program also focuses on improving authentication, data security, multi-cloud provisioning, and provenance tracking, all contributing to the advancement of the TES standard and its ecosystem.

\begin{figure}
    \centering
    \includegraphics[width=\linewidth]{fig5_tel_ap.png}
    \caption{Workflow execution via a TES Gateway allows fine-granular scheduling of individual sequential or parallel workflow steps to suitable TES nodes, e.g. the geographically closest ones, for execution. Integration between the TES Gateway, nodes and clients allows for transparent execution of map / reduce or scatter / gather-type workflow tasks. Due to its support for multiple workflow engines like Nextflow, Snakemake and CWL, TES enables workflow agnostic execution.}
    \label{fig:tel_ap}
\end{figure}

\section{Conclusion and Future Work}
In this paper, we argue that choosing the right compute environment is a challenging and far-reaching decision (\ref{ss:computing_in_academia}). Hybrid and multi-cloud solutions, which are less restrictive as they combine multiple environments (\ref{ss:deployment_models}), are therefore becoming increasingly relevant, provided that adequate abstractions for storage and compute exist on top of the underlying infrastructure.

Additionally, some of the main challenges of computing in research, such as resource scarcity, interoperability, high availability and handling of sensitive data (\ref{s:challenges}), can be addressed by hybrid or multi-cloud architectures (\ref{s:demonstrators}).

As there are multiple ways to implement hybrid and multi-cloud environments, we have explored various approaches ranging from project level (\ref{ss:manual_ap}, \ref{ss:vpn_ap}) to organisation- and federation-wide solutions (\ref{ss:federated_ap}, \ref{ss:wml_ap}, \ref{ss:tel_ap}), demonstrating the flexibility to meet diverse research needs.

By implementing a multitude of multi-cloud approaches of different complexity (\ref{s:demonstrators}), ranging from manual to automatic setups, we showed that multi-cloud approaches are already operational. It primarily remains up to the infrastructures to deploy them and up to the researchers to decide to use these deployments.

We believe that broader adoption of multi-cloud solutions in the future hinges on two critical factors. First, while the diversity of academic and research infrastructures provides a rich ecosystem of computing environments, it also highlights the technological challenge of lacking common service provisioning models, standardisation, and interoperable technologies. Second, the advancement of these systems will depend on robust governance enablers that foster inter-organisational and international cooperation through coordinated policy, sustainable financing, and supportive frameworks. Addressing both the technological and governance challenges is essential to fully realise the potential of hybrid cloud architectures in life science research.

Future work should focus on leveraging the collaborative, as well as the technical and semantic interoperability framework of the European Open Science Cloud (EOSC). Together with the establishment of the first EOSC Nodes and aligned with the EOSC Federation handbook \cite{EOSCFederationHandbooka}, these technical foundations also promise to support the introduction of common governance models for barrier-free interoperability, although this largely remains a political and legal problem. This model can be realised only in close involvement of the already existing ecosystem of research infrastructures such as ELIXIR, which have an indispensable role in coordinating both technological and scientific communities. 

\section{Declaration of competing interest}

The authors declare that they have no known competing financial interests or personal relationships that could have appeared to influence the work reported in this paper.

\section{Acknowledgements}

This work was carried out as part of the Compute Platform Task 2 commissioned service 2022-ECP2 funded by ELIXIR-Europe.
This work was supported by the de.NBI Cloud within the German Network for Bioinformatics Infrastructure (de.NBI) and ELIXIR-DE (Forschungszentrum Jülich and W-de.NBI-001, W-de.NBI-004, W-de.NBI-008, W-de.NBI-010, W-de.NBI-013, W-de.NBI-014, W-de.NBI-016, W-de.NBI-022). Part of this work was carried out with the support of ELIXIR CZ Research Infrastructure (ID LM2023055, MEYS CR).

\section{Data Availability}

Data has been made available where possible and is cited accordingly. Additional data can be made available on an individual basis on request.

\section{Statement on the use of AI}
During the preparation of this work the authors used DeepL Write to improve grammar, sentence structure and style of the original draft manuscript. After using this tool/service, the authors reviewed and edited the content as needed and take full responsibility for the content of the published article. 

%\appendix
%\section{Example Appendix Section}
%\label{app1}


%%       \citet{<label>} ==> Lamport (1994)
%%       \citep{<label>} ==> (Lamport, 1994)
%%

\bibliographystyle{elsarticle-num} 
\bibliography{elixir-compute-hc.bib}

@webpage{ApacheAtlasDataa,
  title = {Apache {{Atlas}} -- {{Data Governance}} and {{Metadata}} Framework for {{Hadoop}}},
  url = {https://atlas.apache.org/#/},
  lastchecked = {2026-01-07},
  year = {2026},
  month = jan
}

@webpage{ApptainerPortableReproduciblea,
  title = {Apptainer - {{Portable}}, {{Reproducible Containers}}},
  url = {https://apptainer.org/},
  lastchecked = {2026-01-07},
  year = {2026},
  month = jan
}

@incollection{bettiviaProvONE2022a,
  title = {{{ProvONE}}},
  booktitle = {Documenting the {{Future}}: {{Navigating Provenance Metadata Standards}}},
  author = {Bettivia, Rhiannon and Cheng, Yi-Yun and Gryk, Michael Robert},
  editor = {Bettivia, Rhiannon and Cheng, Yi-Yun and Gryk, Michael Robert},
  year = {2022},
  series = {Synthesis {{Lectures}} on {{Information Concepts}}, {{Retrieval}}, and {{Services}}},
  pages = {41--56},
  publisher = {Springer International Publishing},
  address = {Cham},
  doi = {10.1007/978-3-031-18700-1_4},
  url = {https://doi.org/10.1007/978-3-031-18700-1_4},
  lastchecked = {2024-02-16},
  isbn = {978-3-031-18700-1},
  langid = {english}
}

@webpage{BiBiGridHybridMultiCloud2024a,
  title = {{{BiBiGrid Hybrid-}} and {{Multi-Cloud Hands-on}}},
  year = {2024},
  month = jan,
  publisher = {German Network for Bioinformatics Infrastructure},
  url = {https://github.com/deNBI/bibigrid_hybrid_cloud_elixir},
  lastchecked = {2026-01-07},
}

@book{commissionRealisingEuropeanOpen2016,
  title = {Realising the {{European}} Open Science Cloud: First Report and Recommendations of the {{Commission}} High Level Expert Group on the {{European}} Open Science Cloud},
  shorttitle = {Realising the {{European}} Open Science Cloud},
  author = {{Directorate-General for Research {and} Innovation (European Commission)}},
  year = {2016},
  publisher = {Publications Office of the European Union},
  address = {LU},
  url = {https://data.europa.eu/doi/10.2777/940154},
  lastchecked = {2023-11-07},
  isbn = {978-92-79-61762-1},
  langid = {english}
}

@article{crusoeMethodsIncludedStandardizing2022a,
  title = {Methods Included: Standardizing Computational Reuse and Portability with the {{Common Workflow Language}}},
  shorttitle = {Methods Included},
  author = {Crusoe, Michael R. and Abeln, Sanne and Iosup, Alexandru and Amstutz, Peter and Chilton, John and Tijani{\'c}, Neboj{\v s}a and M{\'e}nager, Herv{\'e} and {Soiland-Reyes}, Stian and Gavrilovi{\'c}, Bogdan and Goble, Carole and Community, The CWL},
  year = {2022},
  journal = {Communications of the ACM},
  volume = {65},
  number = {6},
  pages = {54--63},
  issn = {0001-0782},
  doi = {10.1145/3486897},
  url = {https://dl.acm.org/doi/10.1145/3486897},
  lastchecked = {2023-11-07}
}

@article{daveigaleprevostBioContainersOpensourceCommunitydriven2017a,
  title = {{{BioContainers}}: An Open-Source and Community-Driven Framework for Software Standardization},
  shorttitle = {{{BioContainers}}},
  author = {{da Veiga Leprevost}, Felipe and Gr{\"u}ning, Bj{\"o}rn A. and Alves Aflitos, Saulo and R{\"o}st, Hannes L. and Uszkoreit, Julian and Barsnes, Harald and Vaudel, Marc and Moreno, Pablo and Gatto, Laurent and Weber, Jonas and Bai, Mingze and Jimenez, Rafael C. and Sachsenberg, Timo and Pfeuffer, Julianus and Vera Alvarez, Roberto and Griss, Johannes and Nesvizhskii, Alexey I. and {Perez-Riverol}, Yasset},
  year = {2017},
  month = aug,
  journal = {Bioinformatics (Oxford, England)},
  volume = {33},
  number = {16},
  pages = {2580--2582},
  issn = {1367-4811},
  doi = {10.1093/bioinformatics/btx192},
  langid = {english},
  pmcid = {PMC5870671},
  pmid = {28379341}
}

@article{ditommasoNextflowEnablesReproducible2017b,
  title = {Nextflow Enables Reproducible Computational Workflows},
  author = {Di Tommaso, Paolo and Chatzou, Maria and Floden, Evan W. and Barja, Pablo Prieto and Palumbo, Emilio and Notredame, Cedric},
  year = {2017},
  month = apr,
  journal = {Nature Biotechnology},
  volume = {35},
  number = {4},
  pages = {316--319},
  issn = {1546-1696},
  doi = {10.1038/nbt.3820},
  url = {https://www.nature.com/articles/nbt.3820},
  lastchecked = {2024-08-14},
  copyright = {2017 Springer Nature America, Inc.},
  langid = {english}
}

@webpage{DnsmasqNetworkServicesa,
  title = {Dnsmasq - Network Services for Small Networks.},
  url = {https://thekelleys.org.uk/dnsmasq/doc.html},
  lastchecked = {2026-01-07},
  year = {2026},
  month = jan
}

@webpage{DockerAcceleratedContainer2022a,
  title = {Docker: {{Accelerated Container Application Development}}},
  shorttitle = {Docker},
  year = {2022},
  month = may,
  url = {https://www.docker.com/},
  lastchecked = {2026-01-07},
  langid = {american}
}

@webpage{donenfeldWireGuardFastModerna,
  title = {{{WireGuard}}: Fast, Modern, Secure {{VPN}} Tunnel},
  shorttitle = {{{WireGuard}}},
  author = {Donenfeld, Jason A.},
  url = {https://www.wireguard.com/},
  lastchecked = {2026-01-07},
  langid = {english},
  year = {2024},
  month = feb
}

@article{duyxScientificCitationsFavor2017a,
  title = {Scientific Citations Favor Positive Results: A Systematic Review and Meta-Analysis},
  shorttitle = {Scientific Citations Favor Positive Results},
  author = {Duyx, Bram and Urlings, Miriam J. E. and Swaen, Gerard M. H. and Bouter, Lex M. and Zeegers, Maurice P.},
  year = {2017},
  month = aug,
  journal = {Journal of Clinical Epidemiology},
  volume = {88},
  pages = {92--101},
  issn = {1878-5921},
  doi = {10.1016/j.jclinepi.2017.06.002},
  langid = {english},
  pmid = {28603008}
}

@webpage{EGIAdvancedComputinga,
  title = {{{EGI}}: {{Advanced Computing}} for a {{Data-Driven Future}}},
  journal = {EGI Foundation},
  url = {https://www.egi.eu/},
  langid = {english},
  lastchecked = {2026-01-07},
  year = {2026},
  month = jan
}

@webpage{EOSCFederationHandbooka,
  title = {{{EOSC Federation Handbook}}},
  journal = {EOSC Association},
  author = {{EOSC Association}},
  url = {https://eosc.eu/eosc-federation-handbook/},
  langid = {british},
  lastchecked = {2026-01-07},
  year = {2026},
  month = jan
}

@webpage{EOSCTITANProjecta,
  title = {{{EOSC TITAN Project}}},
  url = {https://titan-eosc.eu/},
  lastchecked = {2026-01-07},
  year = {2026},
  month = jan
}

@webpage{EuropeanGenomicDatab,
  title = {European {{Genomic Data Infrastructure}} ({{GDI}}) Project},
  url = {https://gdi.onemilliongenomes.eu/},
  lastchecked = {2026-01-07},
  langid = {english},
  year = {2024},
  month = aug
}

@webpage{EuropeanHealthData2024a,
  title = {European {{Health Data Space}} - {{European Commission}}},
  year = {2024},
  month = apr,
  url = {https://health.ec.europa.eu/ehealth-digital-health-and-care/european-health-data-space_en},
  lastchecked = {2026-01-07},
  langid = {english}
}

@webpage{EuropeanNetworkTrusteda,
  title = {European Network of Trusted Research Environments ({{EOSC-ENTRUST}}) Project},
  journal = {EOSC-ENTRUST},
  url = {https://eosc-entrust.eu/},
  lastchecked = {2026-01-07},
  langid = {english},
  month = feb,
  year = {2025}
}

@webpage{ExecutorsNextflowDocumentationa,
  title = {Executors --- {{Nextflow}} Documentation},
  url = {https://www.nextflow.io/docs/latest/executor.html},
  lastchecked = {2026-01-07},
  year = {2026},
  month = jan
}

@webpage{GEANTAnnualReporta,
  title = {{{G{\'E}ANT}} -- {{Annual Report}} 2022},
  url = {https://ar2022.geant.org/},
  lastchecked = {2026-01-07},
  year = {2022}
}

@webpage{GlobalAlianceGenomicsa,
  title = {Global {{Aliance}} for {{Genomics}} and {{Health}} - {{GH4GA}}},
  url = {https://www.ga4gh.org/},
  lastchecked = {2026-01-07},
  langid = {british},
  year = {2025},
  month = feb
}

@webpage{GreenCloudGreen2023a,
  title = {Green Cloud and Green Data Centres {\textbackslash}textbar {{Shaping Europe}}'s Digital Future},
  year = {2023},
  month = dec,
  url = {https://digital-strategy.ec.europa.eu/en/policies/green-cloud},
  lastchecked = {2026-01-07},
  langid = {english}
}

@article{hoffmannEmbeddingDeNBICloud2023a,
  title = {Embedding the de.{{NBI Cloud}} in the {{National Research Data Infrastructure Activities}}},
  author = {Hoffmann, Nils and Maus, Irena and Beier, Sebastian and Belmann, Peter and Kr{\"u}ger, Jan and Tauch, Andreas and Goesmann, Alexander and Eils, Roland and Bork, Peer and Kohlbacher, Oliver and Kummer, Ursula and Backofen, Rolf and Buchhalter, Ivo and Sczyrba, Alexander},
  year = {2023},
  month = sep,
  journal = {Proceedings of the Conference on Research Data Infrastructure},
  volume = {1},
  issn = {2941-296X},
  doi = {10.52825/cordi.v1i.387},
  url = {https://www.tib-op.org/ojs/index.php/CoRDI/article/view/387},
  lastchecked = {2023-12-22},
  copyright = {Copyright (c) 2023 Nils Hoffmann, Irena Maus, Sebastian Beier, Peter Belmann, Jan Kr{\"u}ger, Andreas Tauch, Alexander Goesmann, Roland Eils, Peer Bork, Oliver Kohlbacher, Ursula Kummer, Rolf Backofen, Ivo Buchhalter, Alexander Sczyrba},
  langid = {english}
}

@techreport{infrastructuresESFRIRoadmap20212021,
  title = {{{ESFRI Roadmap}} 2021 - Strategy Report on Research Infrastructures},
  author = {{European Strategy Forum on Research Infrastructures}},
  institution = {{European Strategy Forum on Research Infrastructures}},
  year = {2021},
  month = nov,
  pages = {243},
  url = {https://roadmap2021.esfri.eu/media/1295/esfri-roadmap-2021.pdf},
  lastchecked = {2026-01-07}
}

@article{ioannidisHowMakeMore2014a,
  title = {How to Make More Published Research True},
  author = {Ioannidis, John P. A.},
  year = {2014},
  month = oct,
  journal = {PLoS medicine},
  volume = {11},
  number = {10},
  pages = {e1001747},
  issn = {1549-1676},
  doi = {10.1371/journal.pmed.1001747},
  langid = {english},
  pmcid = {PMC4204808},
  pmid = {25334033}
}

@article{ioannidisWhyMostPublished2005a,
  title = {Why Most Published Research Findings Are False},
  author = {Ioannidis, John P. A.},
  year = {2005},
  month = aug,
  journal = {PLoS medicine},
  volume = {2},
  number = {8},
  pages = {e124},
  issn = {1549-1676},
  doi = {10.1371/journal.pmed.0020124},
  langid = {english},
  pmcid = {PMC1182327},
  pmid = {16060722}
}

@webpage{iso/iec22123-1:2023eInformationTechnologyCloud2023a,
  title = {Information Technology --- {{Cloud}} Computing --- {{Part}} 1: {{Vocabulary}}},
  shorttitle = {{{ISO}}/{{IEC}} 22123-1},
  author = {{ISO/IEC 22123-1:2023(E)}},
  year = {2023},
  month = feb,
  journal = {ISO},
  url = {https://www.iso.org/standard/82758.html},
  lastchecked = {2026-01-07},
  langid = {english}
}

@webpage{iso/iec2382:2015enISOIEC23822015a,
  title = {{{ISO}}/{{IEC}} 2382:2015(En), {{Information}} Technology --- {{Vocabulary}}},
  author = {ISO/IEC 2382:2015(en)},
  year = {2015},
  url = {https://www.iso.org/obp/ui/#iso:std:iso-iec:2382:ed-1:v2:en},
  lastchecked = {2026-01-07},
}

@article{kanitzGA4GHTaskExecution2024a,
  title = {The {{GA4GH Task Execution API}}: {{Enabling Easy Multi Cloud Task Execution}}},
  shorttitle = {The {{GA4GH Task Execution API}}},
  author = {Kanitz, Alexander and McLoughlin, Matthew H. and Beckman, Liam and Malladi, Venkat S. and Ellrott, Kyle},
  year = {2024},
  journal = {Computing in Science \& Engineering},
  pages = {1--16},
  issn = {1558-366X},
  doi = {10.1109/MCSE.2024.3414994},
  url = {https://ieeexplore.ieee.org/document/10564576},
  lastchecked = {2024-08-14}
}

@article{kosterSnakemakeScalableBioinformatics2012,
  title = {Snakemake--a Scalable Bioinformatics Workflow Engine},
  author = {K{\"o}ster, Johannes and Rahmann, Sven},
  year = {2012},
  month = oct,
  journal = {Bioinformatics (Oxford, England)},
  volume = {28},
  number = {19},
  pages = {2520--2522},
  issn = {1367-4811},
  doi = {10.1093/bioinformatics/bts480},
  langid = {english},
  pmid = {22908215}
}

@webpage{Kubernetes2023a,
  title = {Kubernetes},
  year = {2023},
  month = jul,
  journal = {Production-Grade Container Orchestration},
  url = {https://kubernetes.io/},
  lastchecked = {2026-01-07},
  langid = {english}
}

@techreport{mellNISTDefinitionCloud2011a,
  title = {The {{NIST Definition}} of {{Cloud Computing}}},
  author = {Mell, Peter and Grance, Tim},
  year = {2011},
  month = sep,
  number = {NIST Special Publication (SP) 800-145},
  institution = {{National Institute of Standards and Technology}},
  doi = {10.6028/NIST.SP.800-145},
  url = {https://csrc.nist.gov/pubs/sp/800/145/final},
  lastchecked = {2023-11-07},
  langid = {english}
}

@webpage{molderSustainableDataAnalysis2021a,
  title = {Sustainable Data Analysis with {{Snakemake}}},
  author = {M{\"o}lder, Felix and Jablonski, Kim Philipp and Letcher, Brice and Hall, Michael B. and {Tomkins-Tinch}, Christopher H. and Sochat, Vanessa and Forster, Jan and Lee, Soohyun and Twardziok, Sven O. and Kanitz, Alexander and Wilm, Andreas and Holtgrewe, Manuel and Rahmann, Sven and Nahnsen, Sven and K{\"o}ster, Johannes},
  year = {2021},
  month = apr,
  publisher = {F1000Research},
  doi = {10.12688/f1000research.29032.2},
  url = {https://f1000research.com/articles/10-33},
  lastchecked = {2026-01-07},
  copyright = {http://creativecommons.org/licenses/by/4.0/},
  langid = {english}
}

@webpage{NFCoreSarekPipelinea,
  title = {{{NF-Core}}: {{Sarek Pipeline}}},
  shorttitle = {Sarek},
  url = {https://nf-co.re/sarek/3.5.1.html},
  lastchecked = {2026-01-07},
  langid = {english},
  year = {2025},
  month = feb
}

@webpage{OpenPBSOpenSourcea,
  title = {{{OpenPBS Open Source Project}}},
  url = {https://www.openpbs.org/},
  lastchecked = {2026-01-07},
  year = {2026},
  month = jan
}

@webpage{OpenStack2023a,
  title = {{{OpenStack}}},
  year = {2023},
  month = jul,
  journal = {OpenStack},
  url = {https://www.openstack.org/},
  lastchecked = {2026-01-07},
  langid = {english}
}

@webpage{schneider-lunitzElixircloudaaiDemoteshybridcloud1002025,
  title = {Elixir-Cloud-Aai/Demo-Tes-Hybrid-Cloud: 1.0.0},
  shorttitle = {Elixir-Cloud-Aai/Demo-Tes-Hybrid-Cloud},
  author = {{Schneider-Lunitz}, Valentin and Kanitz, Alex},
  year = {2025},
  month = sep,
  doi = {10.5281/zenodo.17038656},
  url = {https://zenodo.org/records/17038656},
  lastchecked = {2026-01-07},
  howpublished = {Zenodo}
}

@article{sheffieldBiomedicalCloudPlatforms2022a,
  title = {From Biomedical Cloud Platforms to Microservices: Next Steps in {{FAIR}} Data and Analysis},
  shorttitle = {From Biomedical Cloud Platforms to Microservices},
  author = {Sheffield, Nathan C. and Bonazzi, Vivien R. and Bourne, Philip E. and Burdett, Tony and Clark, Timothy and Grossman, Robert L. and Spjuth, Ola and Yates, Andrew D.},
  year = {2022},
  month = sep,
  journal = {Scientific Data},
  volume = {9},
  number = {1},
  pages = {553},
  issn = {2052-4463},
  doi = {10.1038/s41597-022-01619-5},
  url = {https://www.nature.com/articles/s41597-022-01619-5},
  lastchecked = {2024-02-16},
  copyright = {2022 The Author(s)},
  langid = {english}
}

@webpage{SlurmWorkloadManager2023a,
  title = {Slurm {{Workload Manager}}},
  year = {2023},
  month = jul,
  journal = {Slurm Workload Manager - Documentation},
  url = {https://slurm.schedmd.com/},
  lastchecked = {2026-01-07},
}

@webpage{spisakovaCERITSCK8shpc2022,
  title = {{{CERIT-SC}}/K8shpc},
  author = {Spi{\v s}akov{\'a}, Vikt{\'o}ria and Hejtm{\'a}nek, Luk{\'a}{\v s}},
  year = {2022},
  month = oct,
  url = {https://github.com/CERIT-SC/k8shpc},
  lastchecked = {2026-01-07},
  copyright = {GPL-3.0},
  howpublished = {CERIT Scientific Cloud}
}

@webpage{stiensmeierBiBiServBibigrid312025,
  title = {{{BiBiServ}}/Bibigrid: 3.1},
  shorttitle = {{{BiBiServ}}/Bibigrid},
  author = {Stiensmeier, Xaver and Kr{\"u}ger, Jan and Dilger, Tim and Friedrichs, Marcel and Henke, Christian and Walender, Alex and Weinholz, David and Sczyrba, Alexander and Belmann, Peter and Kensche, Philip Reiner and Osterholz, Benedikt},
  year = {2025},
  month = sep,
  doi = {10.5281/zenodo.17192507},
  url = {https://zenodo.org/records/17192507},
  lastchecked = {2026-01-07},
  howpublished = {Zenodo}
}

@article{tiokhinHonestSignalingAcademic2021a,
  title = {Honest Signaling in Academic Publishing},
  author = {Tiokhin, Leonid and Panchanathan, Karthik and Lakens, Daniel and Vazire, Simine and Morgan, Thomas and Zollman, Kevin},
  year = {2021},
  journal = {PloS One},
  volume = {16},
  number = {2},
  pages = {e0246675},
  issn = {1932-6203},
  doi = {10.1371/journal.pone.0246675},
  langid = {english},
  pmcid = {PMC7901761},
  pmid = {33621261}
}

@article{wilkinsonFAIRGuidingPrinciples2016a,
  title = {The {{FAIR Guiding Principles}} for Scientific Data Management and Stewardship},
  author = {Wilkinson, Mark D. and Dumontier, Michel and Aalbersberg, I. Jsbrand Jan and Appleton, Gabrielle and Axton, Myles and Baak, Arie and Blomberg, Niklas and Boiten, Jan-Willem and {da Silva Santos}, Luiz Bonino and Bourne, Philip E. and Bouwman, Jildau and Brookes, Anthony J. and Clark, Tim and Crosas, Merc{\`e} and Dillo, Ingrid and Dumon, Olivier and Edmunds, Scott and Evelo, Chris T. and Finkers, Richard and {Gonzalez-Beltran}, Alejandra and Gray, Alasdair J. G. and Groth, Paul and Goble, Carole and Grethe, Jeffrey S. and Heringa, Jaap and {'t Hoen}, Peter A. C. and Hooft, Rob and Kuhn, Tobias and Kok, Ruben and Kok, Joost and Lusher, Scott J. and Martone, Maryann E. and Mons, Albert and Packer, Abel L. and Persson, Bengt and {Rocca-Serra}, Philippe and Roos, Marco and {van Schaik}, Rene and Sansone, Susanna-Assunta and Schultes, Erik and Sengstag, Thierry and Slater, Ted and Strawn, George and Swertz, Morris A. and Thompson, Mark and {van der Lei}, Johan and {van Mulligen}, Erik and Velterop, Jan and Waagmeester, Andra and Wittenburg, Peter and Wolstencroft, Katherine and Zhao, Jun and Mons, Barend},
  year = {2016},
  month = mar,
  journal = {Scientific Data},
  volume = {3},
  pages = {160018},
  issn = {2052-4463},
  doi = {10.1038/sdata.2016.18},
  langid = {english},
  pmcid = {PMC4792175},
  pmid = {26978244}
}

@article{wrattenReproducibleScalableShareable2021a,
  title = {Reproducible, Scalable, and Shareable Analysis Pipelines with Bioinformatics Workflow Managers},
  author = {Wratten, Laura and Wilm, Andreas and G{\"o}ke, Jonathan},
  year = {2021},
  month = oct,
  journal = {Nature Methods},
  volume = {18},
  number = {10},
  pages = {1161--1168},
  issn = {1548-7105},
  doi = {10.1038/s41592-021-01254-9},
  langid = {english},
  pmid = {34556866}
}

@webpage{yuanCovidsequenceanalysisworkflow2023,
  title = {Covid-Sequence-Analysis-Workflow},
  author = {Yuan, David},
  year = {2023},
  publisher = {WorkflowHub},
  doi = {10.48546/WORKFLOWHUB.WORKFLOW.664.1},
  url = {https://workflowhub.eu/workflows/664?version=1},
  lastchecked = {2026-01-07},
  archiveprefix = {WorkflowHub}
}

@article{ziemannFivePillarsComputational2023a,
  title = {The Five Pillars of Computational Reproducibility: Bioinformatics and Beyond},
  shorttitle = {The Five Pillars of Computational Reproducibility},
  author = {Ziemann, Mark and Poulain, Pierre and Bora, Anusuiya},
  year = {2023},
  month = oct,
  journal = {Briefings in Bioinformatics},
  volume = {24},
  number = {6},
  pages = {bbad375},
  issn = {1467-5463},
  doi = {10.1093/bib/bbad375},
  url = {https://www.ncbi.nlm.nih.gov/pmc/articles/PMC10591307/},
  lastchecked = {2023-11-09},
  pmcid = {PMC10591307},
  pmid = {37870287}
}

@webpage{gdpr2016,
  institution = {European Parliament and of the Council},
  year = {2016},
  title = {General Data Protection Regulation {{(EU)}}},
  type = {Regulation},
  number = {2016/679},
  official_journal = {OJ L 119},
  date = {4.5.2016},
  pages = {1-88},
  lastchecked = {2026-01-07},
  url = {https://eur-lex.europa.eu/eli/reg/2016/679/oj/eng}
}



%\begin{thebibliography}{00}


%\bibitem[Lamport(1994)]{lamport94}
%  \textit{\LaTeX: a document preparation system},
%  Addison Wesley, Massachusetts,

%\end{thebibliography}
\end{document}